\documentclass[11pt]{article}

\usepackage{graphicx}

\usepackage{fancyheadings}

\begin{document}

\title{Beyond the Standard Litany: LOSP and Higgs Portals; Lattice Lattice Gauge Theory\footnote{Invited talk at the awards session, EPS-HEPP meeting, Krakow, July 2009.  To be published in the Proceedings.}}

\author{Frank Wilczek \\
\small\it Center for Theoretical Physics\\[-1ex]
\small \it Department of Physics\\[-1ex]
\small \it Massachusetts Institute of Technology\\[-1ex]
\small\it Cambridge, MA 02139}
\date{}
\maketitle

\thispagestyle{fancy}


\newpage
\frenchspacing

I'd like to thank the organizers for inviting me to embellish this EPS Awards session; and I'd like especially to congratulate the Gargamelle collaboration on their long anticipated, amply deserved recognition.  Gargamelle's discovery of neutral currents was the first great discovery I witnessed occurring in real time, so to speak, as a young graduate student.  I remember the widespread skepticism they faced, based on earlier null results and on the difficulty of ruling out backgrounds.   They were courageous, and they were right (as all now agree).    Their discovery was the first, crucial breakthrough, on the experimental side, toward today's electroweak standard model.   Well done!

The organizers didn't give me a specific assignment, just a chunk of time.   My first thought, and presumably the default expectation, was that I'd talk in a general way about the grandeur of high energy physics and the exciting future we anticipate, especially with LHC coming on line.  But an unfortunate byproduct of the delays at LHC and the slow pace of other experiments is that I've already done that sort of thing many times, as have many other conference embellishers, and the ritual is getting stale.  So I'll speed through the litany, and then indulge myself by advertising some ideas I've been fond of, some for quite a while, that I think are interesting, promising, and not as well known as they should be.

\section{The Litany \cite{fwLitany}}

{\bf Where is fundamental physics today?}

\begin{itemize}
\item The standard model is in great shape, empirically.  Theory and experiment agree on many precisely predicted and accurately measured quantities, leaving little room for radiative corrections from physics beyond the standard model.   
\item Flavor physics, including an impressive phenomenology of heavy-meson decays and $CP$ violation, is well described by the Cabibbo-Kobayashi-Maskawa framework.   In particular, there are very strong constraints on additional sources of CP violation, and on additional flavor-changing processes.   
\item Despite the economy, scope, and overwhelming empirical success of the standard model, which would be difficult to overstate, in its present form it has serious shortcomings.   
\end{itemize}

\bigskip

{\bf What are the shortcomings of the standard model?}

\bigskip

\begin{itemize}
\item In the standard model as it comes, the pattern of groups, representations, and (especially) hypercharges is a patchwork. 
\item In the standard model as it comes a great discovery of recent years, the existence of small non-zero neutrino masses, appears gratuitous.   One can accommodate those masses using non-renormalizable, mass dimension 5 operators, but even then the fact that masses of neutrinos are far far smaller than masses of their charged lepton cousins lacks context.
\item In the standard model as it comes there is no evident link between the standard model and the remaining major force of Nature, i.e. gravity.
\item In the standard model as it comes there is no evident accounting for the bulk of the mass in the universe, i.e. the dark matter and dark energy.
\item In the standard model as it comes there is no explanation for the empirical smallness of the $P$ and $T$ violating $\theta$ term.  
\item Masses and mixing angles among quarks and leptons are not constrained by any deep principles, and this sector of the standard model brings in a plethora of unworthy ``fundamental'' constants.
\item The triplication of families is gratuitous.
\end{itemize}

\bigskip

{\bf What are the great lessons of the standard model? }

\bigskip

\begin{itemize}
\item The gauge principle translates mathematical symmetry into physical law.
\item The observed interaction symmetries and fermion representations fit  into a larger, unifying symmetry and unifying representation very economically.
\item Gauge symmetry can be spontaneously broken.
\item Coupling constants evolve with distance, or energy. 
\end{itemize}

\bigskip

{\bf How do those lessons suggest theories to transcend the standard model?}

\bigskip

By drawing on these lessons we can address the first deficiency of the standard model convincingly.   
  
{}The quantum numbers within one family of fermions click together nicely into a spinor representation of $SO(10)$\footnote{There are other possibilities for unifying symmetries, but the successful ones are fairly minor variants of this scheme.}.   The hypercharge assignments, which within the standard model appeared as scraggly orphan ugly ducklings, are then revealed as nonabelian swans, joining their color and weak isospin kindred.   
That the observed quantum numbers of quarks and leptons permit such unification is quite a striking fact: If the lopsided multiplets and peculiar hypercharges of the observed fermions were lopsided or peculiar in some other way, it would be impossible to assemble them into representations of a reasonably small symmetry group.   ($SO(10)$ is one of the very smallest simple groups that accommodates $SU(3)\times SU(2)\times U(1)$.)   The unifying symmetry must be broken, of course.   But broken symmetries have consequences, both for multiplet structure -- which was our point of departure -- and for dynamics.   

The dynamics of gauge theory allows just one overall coupling per simple group; and therefore a unifying symmetry such as $SO(10)$ requires equality of the coupling parameters of its putative sub-theories $SU(3)$, $SU(2)$, and (properly normalized) $U(1)$.   The observed gauge couplings are far from equal.  But couplings evolve; that is our fourth lesson.   If the unified symmetry is spontaneously broken, it must be restored at high energy (or, equivalently, short distance), so the evolution of the couplings toward high energy must bring them to equality.   Coupling evolution is an effect of vacuum polarization, and depends on the field content of the theory:  As we evolve toward ever higher energies, we must include the effect of additional fields, as the suppressive effect of their mass is overcome by the energy.   Given a hypothesis about the field content, and assuming that we can extrapolate toward high energies using weakly coupled quantum field theory\footnote{This is a self-consistent assumption, which however could be upset by bringing in non-perturbative, strongly-coupled sectors, or opening up extra dimensions or string degrees of freedom.}, we can track the evolution, and calculate whether unifying equality ensues.       

By now, famously: If we assume the known content of the standard model, including the Higgs field, then this unification does not quite work out, quantitatively.   But if we include the fields needed to implement (broken) supersymmetry, with superpartner masses in the range $\sim 10^2 -  10^4$ GeV, then the calculation works much better.    Unification takes place at an energy scale $\sim 10^{16}$ GeV.    The unification will continue to work if we include additional complete $SO(10)$ multiplets with intermediate masses, though the scale will be modified.   

In this way, by addressing the first shortcoming of the standard model, we are led to a wonderful expectation, that superpartners should be accessible to the LHC.   As a bonus, we find that several other shortcomings have also been addressed.   Small but non-zero neutrino masses are generated naturally, by the seesaw mechanism.   $SO(10)$ gives us the $SU(3)\times SU(2) \times U(1)$ singlet ``right-handed neutrino'' we need, and the unification scale motivates its required large mass.   The enormous energy scale for unification, which emerges from the phenomenology of particle physics, is close to the Planck scale of gravity.  This means that the powers of {\it all four\/} basic interactions approach equality.  This is a most remarkable result, since at practically accessible energies and distances gravity is {\it absurdly\/} weak compared to the other basic interactions among fundamental particles.   Low-energy supersymmetry also, in many implementations, produces promising candidates to provide the astronomers' dark matter.   So by relieving the first shortcoming of the standard model on our list we seem to make progress on the subsequent three, as well.   

Unfortunately the final three shortcomings on our list are barely touched by all this.   There is one deep connection, however.  The problem of understanding accurate $P$ and $T$ conservation in the strong interaction -- in other words, the smallness of the $\theta$ term -- is most convincingly addressed by postulating an additional (asymptotic) global symmetry, Peccei-Quinn (PQ) symmetry.    PQ symmetry must be spontaneously broken, and that breaking is accompanied by the emergence of a very light, very weakly interacting pseudoscalar particle, the {\it axion}.   Both the mass and the coupling strength of the axion are inversely proportional to the mass scale $F$ at which Peccei-Quinn symmetry is spontaneously broken.    Only if $F$ is extremely large -- $F \geq 10^{10}$ GeV -- is the existence of the axion compatible with experience.    Had unification not taught us that such large scales are already implicit in standard model physics, it might have seemed quite desperate to invoke them here.    But with unification as the background, it hardly raises an eyebrow.   Axions also provide a good candidate for the astronomers' dark matter.

\section{LOSP and Higgs Portals}

Particles or whole sectors that are singlets under $SU(3)\times SU(2) \times U(1)$ start out half-way towards ``explaining'' their non-observation to date; that is, they have none of the canonical, well-characterized (and not particularly feeble) interactions that dominate the behavior of the particles we know about and can access readily.   The existence of such singlet particles has been suggested many times, with various motivations.    Right-handed neutrinos, axions, and of course gravitons are examples that arose even in our standard litany; other examples include hypothetical scalar dark matter protected by discrete symmetries, singlet additions to the minimal supersymmetric implementation of the standard model (creating the NMSSM from the MSSM), which may have phenomenological advantages, and many others.    

How can we test such speculations?   Each proposal has a specific story of its own, but there are a couple of robust observations, based on general principles, that apply to many of them, and to others that might arise in the future.   Both are based on the idea that certain extremely well-motivated (but as yet unobserved) ``ordinary'' particles have particularly natural ways to communicate with singlet sectors.   In this sense, they serve as portals into hidden worlds.

\subsection{LOSP Portal \cite{LOSPPortal}}

All the particles of the standard model have $R$-parity 1, where 
\begin{equation}
R = (-1)^{3B+L+2S} \nonumber
\end{equation}
with $B$ baryon number, $L$ lepton number, and $S$ spin.  Their superpartners, therefore, will have $R = -1$.   Since $B$, $L$, and $S$ are excellent quantum numbers, we should expect that $R$ parity is very accurately conserved.   This implies that the lightest $R$-parity odd particle -- the lightest superparticle, commonly called the LSP -- will be highly stable.    

The superpartners of standard model particles, if they are not too heavy, will be produced with small but not minuscule rates at the LHC, through processes related to standard model processes by supersymmetry.   So for instance since two gluons can scatter, they can also annihilate, with essentially the same coupling strength, into two gluinos.    Similarly, kinematically allowed decays involving superpartners of standard model particles will occur at typical particle-physics rates.   

However it is easy to imagine that the lightest standard model partner, which it's convenient to call the LOSP (for Lightest Ordinary Supersymmetric Particle), is not the lightest supersymmetric particle.  Two particularly plausible possibilities are that the true LSP is the goldstino\footnote{When we couple in gravity, the goldstino becomes the helicity $\pm\frac{1}{2}$ component of the gravitino, and acquires mass.  But its couplings are essentially unchanged.}  associated with spontaneous supersymmetry breaking, or the supersymmetric partner of the axion, the axino.   In both cases, the couplings of the LSP are inversely proportional to the symmetry breaking scale (for supersymmetry and PQ symmetry, respectively) and could be quite small.   Then we would have very small cross sections for direct production of the LSP.   Nevertheless its existence opens up dramatic new phenomenological possibilities.   

Since it is the final repository of odd $R$ parity, the LOSP would be an end product of most rapid decay chains of SUSY particles produced at LHC.  It will eventually decay into the LSP, but those decays might be quite slow by ordinary accelerator standards.   So experimenters should be alert to the possibility of decays occurring far from the interaction region.   For example, if the LOSP $\tilde \gamma$ has a large photino component, and the axino $\tilde a$ is the LSP, one could have decays $\tilde \gamma \rightarrow  \tilde a \gamma$, resulting in prompt energetic photons well removed from the interaction region.    

Instability of the LOSP also loosens cosmological constraints.  Most dramatically, it opens up the possibility that the LOSP might be electrically charged.    Then one would have a copiously produced, apparently stable charged particle produced at the LHC.   A truly stable particle of this kind would be a cosmological disaster, but lifetimes in the range $\tau \leq 10 $ sec. or so are cosmologically harmless, and could provide quite a spectacle for the LHC.

\subsection{Higgs Portal \cite{higgsPortal}}

With one exception, all the interactions of the standard model are associated with strictly renormalizable interactions.  In other words, the interactions and kinetic terms are represented by operators of mass dimension 4; or, in still other words, their associated couplings are dimensionless, in units with $\hbar = c =1$.   The Higgs field mass term $\Delta {\cal L} = - \mu^2 \phi^\dagger \phi$, with a coupling of mass dimension 2 and an interaction operator of mass dimension 2, is the exception.   Thus the Higgs field is uniquely open to renormalizable (or superrenormalizable) coupling to $SU(3)\times SU(2) \times U(1)$ singlet fields.  Of course, neither Higgs particles nor $SU(3)\times SU(2) \times U(1)$ singlets have been observed as yet: the former, presumably because they are too heavy for existing accelerators; the latter, because no generating source has been available.   But it's intriguing to speculate, based on this observation, that their discovery might be simultaneous. 

Several theoretical ideas motivate the concept of  ``hidden'' sectors consisting of $SU(3)\times SU(2) \times U(1)$ singlet fields.   Independent of any model, we can simply note that known standard model fields couple to a variable number of the standard model gauge fields, from all three for the left-handed quark fields to just one for the right-handed electron.  (And in the seesaw mechanism of neutrino mass generation, heavy $SU(3)\times SU(2) \times U(1)$ singlet fermion fields play a crucial role.)  Thus there is no evident reason, if we envisage additional product gauge fields, not to imagine that there are fields which transform under the new but not under the familiar gauge symmetries.  Specific models containing fields that could be construed as a forming a (tiny) hidden sector have been considered for phenomenological purposes.  The product $E(8)\times E(8)$ structure, with forms of matter transforming only under one or the other factor, arises in heterotic string theory, and many other string theory constructions also lead to structures of that sort.   

Most discussion of hidden sectors has posited that they are associated with a high mass scale.   That assumption immediately explains why the ``hidden'' sector is in fact hidden, but leaves the challenge -- a form of hierarchy problem -- of understanding why interactions do not pull the mass scale of the visible sector close to that high scale.   On the other hand, it is not inconceivable that the intrinsic scale of the hidden sector is smaller than, or comparable to, that of the visible sector.   In that case coupling of the hidden to the visible sector generally occurs only through fields whose masses are naturally pulled up close to the visible scale (what was a bug in the other direction becomes, read this way, a feature).    In this way we find a simple explanation for why a light ``hidden'' sector could in fact have remained hidden to date.   It need not remain hidden, however, once the Higgs portal opens.

\bigskip

Here I'll discuss just one simple example.   A very simple hidden sector, coupled to ordinary matter only through the Higgs mass term, could implement the attractive idea that fundamental interactions contain no explicit mass scale at all.  Indeed, we can let the hidden sector consist of a confining massless gauge theory that completely commutes with  the standard model (s) $SU(3)_s \times SU(2)_s \times U(1)_s$, in the sense that its quarks are $SU(3)_s \times SU(2)_s \times U(1)_s$ singlets.   The effective theory of the hidden sector will be a sort of $\sigma$ model, and the hidden (h) $\sigma_h$ field will couple to the standard model in the form ${\cal L}_{\rm link} =  \eta \phi^\dagger_s \phi_s \sigma_h^2$.   Linear terms are forbidden by hidden sector chiral symmetry.   Spontaneous chiral symmetry breaking with $\langle \sigma_h \rangle = \kappa$ will then generate an effective mass$^2$ $-\eta \kappa^2$ for $\phi_s$, which, assuming $\eta > 0$, could trigger electroweak symmetry breaking.   In this scenario the ratio between the weak scale and the Planck scale arises from an effect in the hidden sector similar to the one that works in QCD to generate the ratio between the proton mass and the Planck scale.   Namely, an enormous disparity of scales can be required in order for a moderate value of the hidden gauge coupling at the Planck scale to evolve to a large value, and induce chiral symmetry breaking, because running of couplings is logarithmic.   As in QCD, no terribly small (unnatural) numerical quantities need be involved.   Of course, one would still have to understand why the intrinsic mass$^2$ of the Higgs field vanishes, or is subdominant.   

\bigskip

In alternative models there could be not only one but several fields from the $SU(3)_s \times SU(2)_s \times U(1)_s$ singlet sector that mix with the conventional Higgs field.   Then the conventional production rate would be shared among several different states, and conventional signals (apart from missing energy-momentum and downstream interactions) would be further diluted by decays into the hidden sector.  These effects would make the physics of the Higgs sector richer, but more challenging to access and sort out experimentally.

\section{Lattice Lattice Gauge Theory}

\subsection{Mott Transitions}

Let me begin by briefly reviewing for you the Mott transition, which is a fundamental, though in detail still poorly understood, issue in solid state physics.   

Consider a lattice of hydrogen atoms, say simple cubic to be definite, where we imagine the spacing between nuclei to be a freely adjustable parameter.  Band theory tells us we should have a metal, since in the ground state the lowest band will be half full.   But this is a paradoxical result, since when the atoms are far apart we would expect the electrons to be localized, one per atom.   Mott's resolution was to emphasize that band theory starts by neglecting electron interactions.  When we reinstate them, we realize that of course there should be an energy gap, which is (for well separated atoms) basically the energy it takes to ionize $H \rightarrow H^+ + e$, minus the binding energy of the second electron in an $H^-$ ion, i.e. the energy liberated in $H + e \rightarrow H^-$.   As the atoms get close together this argument fails, because the atomic electron wave functions start to overlap, and the energetics change.   Indeed let us start with the opposite limit of delocalized electrons (as in band theory).  The mobile electron gas is very efficient at screening charge, and so the repulsive Coulomb interaction loses its potency.   Thus the free-electron band approximation is justified, self-consistently.  This hand-waving argument can be made somewhat more rigorously, of course, but as far as I know the exact nature of the transition -- or transitions? (see below) -- between the two extreme cases of metal at very small spacings and insulator at large spacings has never been fully elucidated.   (Its degree of universality is far from obvious; see immediately below.)  

The Mott phenomenon was long regarded by most condensed matter physicists as an interesting aberration, but basically a departure from the default, i.e. band theory.    It rose to high prominence in connection with high temperature superconductivity, because the parent materials, at zero doping, are Mott insulators\footnote{There's actually a complication, that these parent materials exhibit antiferromagnetic order which, acting as a background, in itself opens a gap. But according to most estimates that effect is too small to account for the measured gap.}   On the other hand chemists, when they think about the solid state, come at it the other way.   For them separated atoms (or perhaps molecules) and their orbitals are the natural starting point, and metallic behavior is the aberration to be explained.   And the chemists have a point: there are huge classes of materials, the so-called covalently bonded, ionic, and molecular solids -- insulators all -- for which an atomic or molecular starting point is manifestly more appropriate than band theory.   

The chemists' perspective serves to emphasize that the ``Mott transition'' between dynamically independent, localized electrons and simple metallic behavior need not be a one-step affair.    One can imagine different intermediate bonding relationships, presumably involving bigger effective molecules or electrons that travel as pairs or other units, rather than entirely independently.   Anderson's resonating valence bond (RVB) state seems to envision something along these lines.   Investigations of solids under high pressure realize a version of Mott's thought experiment, and in some cases very complicated phase diagrams have emerged.  

\subsection{Sign Problems}

Unfortunately nature does not provide us with a wealth of simple, clearly characterized model experimental systems to guide and test our thinking on these issues.   There are ambitious proposals to use ultracold atoms conditioned by optical lattices as such model systems, and that is an exciting prospect, but it is not easy, and we'll have to wait to see what really can be achieved.   I'd like to propose a different kind of toy model, that I think is also interesting in its own right, that should be amenable to numerical treatment.   

The powerful technique of lattice gauge theory uses importance sampling to estimate the gigantic-dimensional integrals that arise.  It has been used with great success to elucidate the properties of individual hadrons, and also the thermodynamics of QCD at zero chemical potential, but its application to nuclear matter at non-zero chemical potential has not got very far, for it is plagued by the notorious sign problem.  Actually this is a phase problem: as we integrate over field configurations, any observable of interest gets contributions that are complex numbers including relative phases, and the magnitude of the final answer is much smaller than the magnitude of the individual contributions.  (Speaking more properly: the integral of the absolute squares is much larger than the square of the integral.)  In other words there are massive cancellations, and that makes it essentially impossible to extract an accurate answer from imperfect data.     Similar problems arise when one attempts to simulate interesting electron systems.   

There are three basic sources of sign and phase problems.   One is simply Fermi statistics.    In QCD (at zero chemical potential) that problem is bypassed, at considerable cost, by integrating out the fermions.  The resulting integrand is then positive, though much more complicated and no longer local.   The calculations for gauge theories including fermions are considerably more arduous and time-consuming than for theories without fermions, but after many improvements in computer hardware and in algorithms the state of the art has become impressive indeed.    The second problem is non-zero chemical potential.   Its presence renders the integrand complex, which is usually disastrous.   One might try to bypass chemical potential by introducing instead explicit quark (or electron) sources, but if the source world lines are dynamical their entanglings give signs by Fermi statistics.  Worse yet -- this is the third problem -- even if we consider sources fixed in space, effectively eliminating quantum statistics, their world lines introduce phases; indeed, it is complex conjugation that distinguishes a quark from an antiquark world line.   

\subsection{Positive Models and Probes}

The integrands are positive definite, however, for systems of bosons, or fixed sources, in representations with positive characters.   Since the Wilson/Polyakov loop factors that implement the interaction of the sources in a representation $R$ with the gauge field are just the characters $\chi^R(g)$ for the parallel transport $g$ around the loop, insertion of such factors maintains the positivity of the integrand.   In fact, we need not even deal with characters of representations -- any non-negative class function is fair game, for the defining property $f(hgh^{-1}) = f(g)$ of a class function is exactly what's required for gauge invariance of the loop integral.  So lattices of fixed sources based on non-negative class functions provide user-friendly model systems.   The spacing of the source lattice can and, in view of the preceding discussion, should be varied independently of the spacing in the numerical grid; thus it's appropriate to speak of {\it Lattice lattice gauge theory}.    We can vary the dimensionality, the type of lattice, the gauge group(s), and the class function(s) present;  we can also include temperature and (still at a high price) dynamical fermions.   

Class functions can be analyzed into linear combination of characters of irreducible representations, using the orthogonality relations.  Non-negative class functions will have non-zero overlap with the identity, so our sources have some amplitude to represent vacuum quantum numbers.   For appropriate groups and class functions, however, the overlap with other representations can be significantly larger.  Taking a finite group and the class function equal to a Kronecker delta at the identity, for example, we find that the overlap with a given irreducible representation is simply the dimension of that representation.   

Of course, for modeling purposes one need not be committed to gauge theories; there may well be other physically interesting models arising as regular arrays of sources governed by manifestly positive local actions.

\bigskip

\bigskip 

\bigskip

There's a lot more to say about each of the ideas I've touched on here, but I hope I've shown you enough to get you thinking.

\bigskip 

\bigskip

\bigskip

{\it Acknowledgement}: I'd like to thank Uwe-Jens Wiese and Michele Pepe for pointing out a serious error in the original version of the last part of this paper.


\begin{thebibliography}{99}
 
 \bibitem{fwLitany}  See for example F. Wilczek, {\it The Future of Particle Physics as a Natural Science}, Published in ``Critical Problems in
Physics'' in celebration of the 250th Anniversary of Princeton University, November 1996, eds. Fitch, Marlow, and Dementi, Princeton University Press; also in Int. Jour. Mod. Phys. A 13, 863, (1998); also in Magazine of Physics, Science \& Ideas Vol. 1 No. 2) 12-25, (Dec. 1996) hep-ph/9702371.
 
 \bibitem{LOSPPortal}  K. Rajagopal, M. Turner, and F. Wilczek, {\it Cosmological Implications of Axinos}, Nucl. Phys. {\bf B358},
447 (1991) is an early paper on the subject, where the term LOSP was first used.  C. Cheung, Y. Nomura, and J. Thaler  {\it Goldstini}, arXiv:1002.1067 is a recent paper with extensive references. 
 
 \bibitem{higgsPortal}   An early model exemplifying this portal is V. Silveira and A. Zee, Phys. Lett. {\bf B161} 136  (1985).  J. J. van der Bij and J. Gunion pioneered the possible accelerator phenomenology of hidden sectors connected to the Higgs particles, and have written extensively on the subject.  See for example J. J. van der Bij {\it No Higgs at the LHC}, arXiv:0804.3534; S. Chang, R. Dermisek, J. Gunion, and N. Weiner {\it Nonstandard Higgs Boson Decays}, Ann. Rev. Nucl. Part. Sci. {\bf 58} 75-98 (2008) and references therein.    The general connection to low-dimension operators is mentioned in F. Wilczek {\it A Constructive Critique of the Three Standard Systems}, (2004) hep-ph/0401126, Advanced Studies Institute: Physics at LHC-Praha-2003, Prague, CR) Vol 54-2004 A415-427 [MIT- CTP-3465], and further emphasized in B. Patt and F. Wilczek, {\it Higgs-field Portal into Hidden Sectors}, hep-ph/0605188.   J. Wells and collaborators have done important work on the subject; see for example R. Schabinger and J. Wells {\it A Minimal Spontaneously Broken Hidden Sector and Its Impact on Higgs Boson Physics at the Large Hadron Collider}, Phys. Rev. {\bf D72}:093007 (2005) and J. Wells {\it How to Find a Hidden World at the Large Hadron Collider}, in Perspectives on LHC Physics ed. G. Kane and A. Pierce (World Scientific, 2008), arXiv:0803.1243.   


\end{thebibliography}
\end{document}